# Identifying Disruptive Models in the Open-Source LLM Community

Xiaoting Wei[1,2], Lele Kang[1,2], Xuelian Pan[1,2], Jiannan Yang[1,2,*]

[1]Nanjing University, Laboratory of Data Intelligence and Interdisciplinary Innovation, China
[2]School of Information Management, Nanjing University, China
*Corresponding Author: Jiannan Yang (jnyang@nju.edu.cn)

**ABSTRACT**

The rapid growth of open-source large language models (LLMs) has created a complex ecosystem of model inheritance and reuse. However, existing research has focused mainly on descriptive analyses of lineage evolution, with limited attention to identifying which models play a disruptive role in shaping subsequent development. Using metadata from 2,556,240 models on Hugging Face, this study reconstructs a large-scale lineage network and introduces the Model Disruption Index (MDI) to distinguish between models that reinforce existing technological trajectories and those that become new bases for later development. The results show that most models in the open-source LLM community are consolidative rather than disruptive, reflecting a highly concentrated and path-dependent evolutionary structure. Further analyses suggest that disruptive positions are more likely to emerge among large-scale models and through finetuning strategies. Overall, this study provides a new perspective for identifying disruptive models and understanding uneven technological development in open-source LLM ecosystems.

**KEYWORDS**

Open-Source Large Language Models, Hugging Face, Model Lineage Network, Technological Disruption, Scientometrics

**1 INTRODUCTION**

Since 2022, the rapid expansion of open-source large language models (LLMs) has made model development increasingly visible on platforms such as Hugging Face. As more models are released through fine-tuning, quantization, adapter training, or merging, open-source LLM communities have evolved far beyond simple repositories of standalone models. Instead, they have become dynamic ecosystems in which new models are often built on earlier ones, forming clear patterns of inheritance and reuse. In such an environment, an important question is no longer only how many models are being released, but also whether the pattern of innovation in the community tends to be consolidative or disruptive.

Recent studies have begun to examine this ecosystem from the perspectives of model lineage, provenance, and supply-chain dependency (Bommasani et al., 2023; Shang et al., 2026; Wu et al., 2026). These studies have shown that open-source LLM development is deeply shaped by relationships among base models, derivative models, datasets, and applications (Rahman et al., 2025; Tamura et al., 2025). However, most existing work remains largely descriptive and focused on major foundation models. While it can reveal how models are connected, it is less able to show whether a newly released model mainly extends an existing technological path or instead becomes a new basis for later development. This distinction matters because a model's importance in an open-source community is not determined only by visibility or popularity. Some models may attract attention while still remaining part of an established path, whereas others may play a more important role in shaping what comes next.

This problem closely parallels a longstanding concern in scientometrics and the study of technological change. Research on papers and patents has long noted that most innovations tend to build on and reinforce existing trajectories, while only a limited number shift later activity in new directions (Redner, 1998; Valverde et al., 2007; Wu et al., 2019; Park et al., 2023). Disruption-based measures were developed to distinguish between innovations that consolidate prior paths and those that redirect subsequent development (Funk & Owen-Smith, 2017; Wu et al., 2019). Although open-source LLM communities differ from citation systems in important ways, they share a similar relational logic: models build on earlier models, and later models may either continue to follow earlier foundations or increasingly turn to a newer model as their point of departure. This makes it possible to apply ideas from scientometrics into the study of open-source model evolution.

Against this background, this study investigates technological disruption in the open-source LLM community on Hugging Face from a scientometrics network perspective. Using metadata collected from Hugging Face, we construct a large-scale lineage network of derivation relationships and introduce the Model Disruption Index (MDI) to evaluate whether a focal model redirects downstream development away from its predecessor trajectory or primarily consolidates it. Guided by this framework, we address three research questions: RQ1, what are the overall evolutionary characteristics of the open-source LLM lineage network? RQ2, do clearly disruptive models emerge within the community? RQ3, what factors are associated with model disruption? By answering these questions, this study extends existing research in three ways. First, it moves beyond descriptive lineage tracing to capture the evolution of technological trajectories derived from focal models. Second, it adapts disruption-based thinking from papers and patents to open-source model communities. Third, it provides new evidence that openness in model access does not necessarily lead to broadly distributed technological change, but may coexist with strong concentration and path dependence.

## 2 LITERATURE REVIEW
### 2.1 LLM Lineage Evolution and Ecosystem Dependencies

With the rapid proliferation of open-source LLMs, derivation relationships among models have formed a large and evolving ecosystem (Tamura et al., 2025). Existing research has begun to examine this ecosystem mainly from the perspectives of model lineage and model supply chains (Bommasani et al., 2023; Shang et al., 2026; Wu et al., 2026). Regarding lineage evolution, researchers have paid increasing attention to the iterative characteristics of core model families such as GPT, LLaMA, and Qwen, highlighting the importance of identifying how models inherit from and build upon one another (Laufer et al., 2025; Shang et al., 2026; Wu et al., 2025; Wu et al., 2026). For example, Wu et al. (2026) proposed the concept of "LLM DNA" and utilized functional representations to track undocumented model evolutionary relationships. Similarly, Wu et al. (2025) constructed "model fingerprints" based on the gradient responses of LLMs under perturbation to detect similarity and support model family classification and provenance tracing. From a broader ecosystem perspective, Laufer et al. (2025) analyzed the lineage characteristics of 1.86 million models on Hugging Face and showed that open-source models exhibit rapid and directional evolutionary patterns during finetuning diffusion. Shang et al. (2026) further emphasized the value of fine-tuning trajectories for lineage verification, while Tamura et al. (2025) suggested that clarifying model lineage relations can also help predict model performance.

Beyond lineage identification itself, recent studies have also examined how models are embedded in broader dependency structures. Rahman et al. (2025) for instance, proposed HuggingGraph to track the supply chain dependencies among base models, derivative models and datasets. Likewise, Bommasani et al. (2023) introduced Ecosystem Graphs to show how foundation models shape technical and social dependencies across datasets, models and applications. Wang & Huang (2025) argued that open-source software contributors tend to expand the evolutionary scale of derivative models rather than base models. Taken together, these studies have significantly advanced our understanding of how open-source LLM ecosystems are organized and how models are connected through inheritance and dependency relationships.

However, existing research has focused mainly on reconstructing lineage relations, tracing provenance, and mapping ecosystem dependencies. While these studies have substantially improved our understanding of how open-source LLMs are connected, they remain less concerned with directional differences in downstream development. In other words, current work is better at showing how models are related than at identifying whether some models play a more consequential role in shifting later development away from inherited trajectories. This limitation points to the need for an analytical perspective that can move beyond descriptive tracing and more directly evaluate the disruptive significance of individual models within the evolving community.

## 2.2 Disruption Identification of Network Nodes

While existing studies have improved our understanding of model lineage and ecosystem dependencies, they remain less able to identify the key models that reshape subsequent development. This question has been more fully explored in the broader literature on technological change and scientometrics, where scholars have long been concerned with how to distinguish innovations that reinforce existing trajectories from those that shift later development in new directions (Funk & Owen-Smith, 2017; H. Park & Magee, 2019; L. Wu et al., 2019). Prior study (Cecere et al., 2014) has noted that technological evolution is often shaped by "path dependence" where small historical events provide a technology with an initial advantage, generating switching costs that exclude alternatives and "lock-in effects" where increasing returns and self-reinforcing processes trap a system in a dominant outcome that is not necessarily superior. These dynamics inevitably lead to a large-scale accumulation of derivative and incremental innovations. To identify the rare and transformative nodes capable of breaking these structural lock-ins, scholars have proposed a range of quantitative approaches within knowledge networks. For example, Park & Magee (2019) explored quantitative indicators for measuring technological discontinuities within knowledge networks. A particularly influential contribution is the $CD$ index proposed by Funk & Owen-Smith (2017), which captures whether an innovation tends to consolidate an existing knowledge structure or redirect subsequent development away from prior trajectories. Applying this approach, Park et al. (2023) conducted a large-scale measurement of tens of millions of papers and patents, revealing a macro-trend of a pervasive decline in the disruptiveness of global scientific and technological innovation.

Subsequent studies further developed this line of research by proposing alternative variants and examining how disruption should be interpreted. For example, Wu et al. (2019) proposed the Disruptive index, which uses citation counts rather than binary indicators to capture whether subsequent work continues to rely on earlier foundations or instead turns more strongly toward a focal contribution. The core logic of the Disruptive index is that when papers citing a specific article also cite a large proportion of that article's references, the article can be regarded as consolidating the scientific field. Conversely, when future citations of the article do not acknowledge its predecessor literature, the article can be viewed as disrupting the field (Ruan et al., 2021). At the same time, previous studies have also pointed out that disruption-based indicators are sensitive to factors such as citation window length and citation volume (Bornmann & Tekles, 2019a, 2019b; Ruan et al., 2021).

Although these studies provide useful tools for identifying disruptive nodes, their application has been concentrated mainly in paper and patent networks. Much less attention has been paid to whether similar ideas can be used to understand open-source model ecosystems. This is an important gap, because models in the open-source LLM community also develop through visible inheritance relationships, and later models may either continue along earlier technological paths or increasingly build on a newer model. Therefore, inspired by Funk & Owen-Smith (2017) and Wu et al., (2019), this study extends disruption-based thinking to the open-source LLM context and proposes a measure for identifying whether a model primarily consolidates an existing trajectory or becomes a new basis for subsequent development.

# 3 METHODS

## 3.1 Data Collection

This study utilizes the public model API provided by the Hugging Face to collect metadata on open-source models hosted on the platform. As of February 3, 2026, a total of 2,556,240 model records were retrieved. The available metadata covered 28 fields, including model identifiers, creation timestamps, and descriptive tags. Among these, the core metadata used in this paper include the model id, creation time, tags, and card-related metadata, which together provide the basis for identifying inter-model derivation relationships in subsequent analysis. Since the earliest creation time available on the Hugging Face platform is March 2, 2022, at 23:29:04, the temporal scope of the dataset spans from March 2, 2022, to February 3, 2026.

## 3.2 Lineage Network Construction

To construct the LLM lineage network, we identified derivation relationships between models from multiple metadata sources rather than relying on a single field. Specifically, parent-child links were extracted through a phased procedure. We first parsed *base_model:\** patterns from model tags, which served as the highest-priority source for identifying predecessor models. For models not covered by tags, we then searched a sequence of fallback fields, including *config_peft_base_model_name_or_path*, *card_base_model*, and *card_data.base_model*. This phased strategy reduced conflicts across metadata sources and allowed us to recover lineage links even when base-model information was incompletely documented in any single field.

Each extracted relation was represented as a directed edge from a subsequent model to its predecessor model. Based on the available metadata, we categorized derivation relations into four substantive types: (i) finetune, where a base model's parameters are updated with new data to improve performance on specific tasks. (ii) adapter, where small trainable modules are added to a frozen base model to learn new capabilities with minimal computational cost. (iii) quantized, where the numerical precision of the model's weights is reduced to decrease memory usage and accelerate inference. (iv) merge, where the weights of multiple models are mathematically combined into a single architecture without additional training. A small number of remaining cases were retained as unspecific when the parent model could be identified but the exact derivation strategy could not be reliably determined. We removed empty values, self-loops and the edges of unspecific relation before assembling the network. The resulting graph was represented as a directed acyclic graph (DAG), in which nodes denote models and edges denote derivation paths from child models to parent models. Figure 1 provides an illustrative subgraph centered on Qwen/Qwen2.5-7B. In this example, Qwen/Qwen2.5-7B-Instruct is connected to its predecessor Qwen/Qwen2.5-7B through a finetune relation, and bartowski/Qwen2.5-7B-Instruct-GGUF is linked through quantization. The model nvidia/Eagle2.5-8B is linked to multiple predecessor models including google/siglip2-so400m-patch16-512 and Qwen/Qwen2.5-7B-Instruct through merging. This example illustrates how different derivation strategies are encoded within the lineage graph used for subsequent MDI analysis.

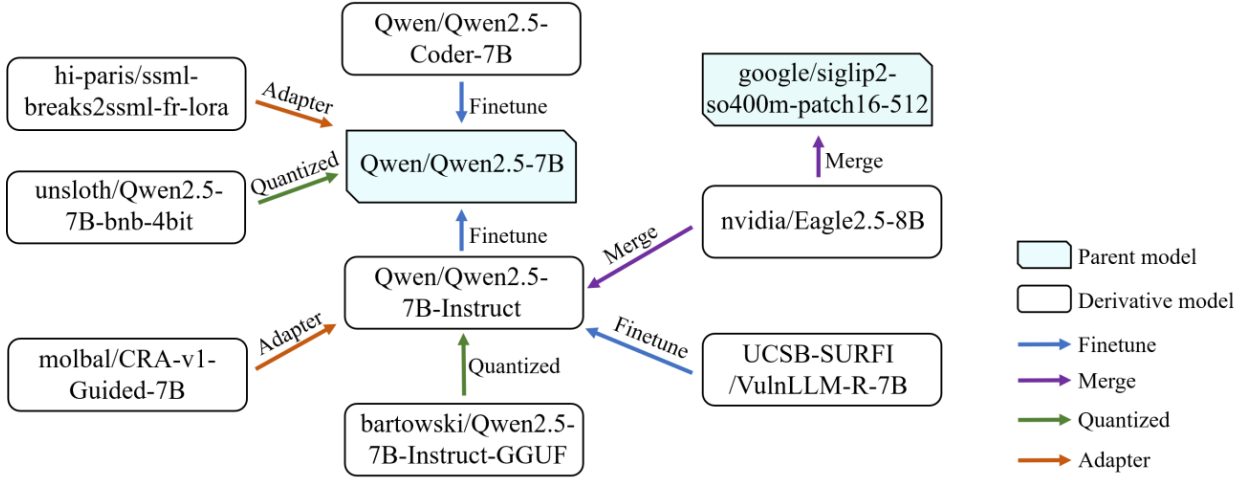

**Figure 1.** An illustrative LLM lineage subgraph centered on "Qwen/Qwen2.5-7B".

### 3.3 Model Disruption Index

To quantify whether a model reinforces an existing technological trajectory or redirects subsequent development away from its predecessors, we adapt the concept of disruption-based measures from scientometrics to the open-source LLM community. In particular, we introduce the Model Disruption Index (MDI) based on the logic underlying the prior disruption measures, including the well-established $CD$ index (Funk & Owen-Smith, 2017) and Disruption index (Wu et al., 2019). In citation-based studies, disruption is assessed by examining whether later papers cite a focal paper alone, continue to co-cite its predecessors, or bypass the focal paper while remaining attached to earlier work. We translate this logic into the LLM context by treating a focal model as analogous to a focal paper, its parent model or parent set as analogous to predecessor references, and subsequent derivative models as analogous to later citing papers. Under this mapping, a model is considered more disruptive when later derivations preferentially build on the focal model rather than continuing to derive from its predecessor models. Conversely, a model is considered more consolidative when subsequent development remains anchored to the parent trajectory.

Formally, for a focal model $i$, let $P_i$ denote its parent model. For merged models, $P_i$ is extended to a parent set. We examine subsequent models released within an observation window of length $t$ after the release of model $i$. In the main analysis, we set $t = 90$ days to capture the model's early downstream uptake in a fast-moving ecosystem while preserving comparability across focal models. Within this window, subsequent models are partitioned into three groups: $X_i(t)$, those that derive from the focal model but not its parent set; $Y_i(t)$, those that derive from both the focal model and at least one parent in the parent set; $Z_i(t)$, those that derive from the parent set but not the focal model. We define the MDI as:

$$MDI_i(t) = \frac{X_i(t) - Z_i(t)}{X_i(t) + Z_i(t) + Y_i(t) + \varepsilon}, \qquad (1)$$

where a small constant $\varepsilon$ is added to the denominator to prevent division by zero when no subsequent derivations are observed within the window.

The design of this formula follows a simple intuition. $X_i(t)$ captures downstream development that shifts to the focal model and no longer remains attached to its predecessor trajectory, whereas $Z_i(t)$ captures development that continues along the predecessor trajectory without adopting the focal model. These two categories therefore represent opposite downstream tendencies, so the numerator is defined as their difference. By contrast, $Y_i(t)$ represents overlapping cases in which the focal model has been adopted but has not fully displaced its predecessors; it is therefore not assigned a directional sign in the numerator, but is retained in the denominator as part of the total observable downstream response. This normalization allows the MDI to range from -1 to 1. Positive values indicate that subsequent derivations preferentially build on the focal model, suggesting a more disruptive position in the lineage network. Negative values indicate that subsequent derivations remain concentrated on the parent trajectory, suggesting a more consolidative position. Values near zero indicate a mixed response, in which the focal model and its predecessor trajectory attract comparable downstream derivations. We illustrate this calculation logic (see Figure 2): within the observation window, subsequent models are grouped according to whether they derive from the focal model only, from both the focal model and its predecessor trajectory, or from the predecessor trajectory only. In the illustrative example, three subsequent models derive from the focal model only, one derives from both, and two derive from the parent trajectory only, yielding a MDI of 0.167 and indicating a slight shift of downstream development toward the focal model.

For most focal models, disruption is evaluated relative to a single parent model. Merged models differ in that they inherit from multiple parent models rather than one. To handle this case, we treat all parent models of a merged focal model as a parent set. A subsequent model is then counted as parent-linked if it derives from at least one model in this set, even if it does not derive from the focal merged model itself. This allows the MDI to assess whether later development is redirected toward the merged focal model or whether it continues to remain attached to one or more of its predecessor trajectories.

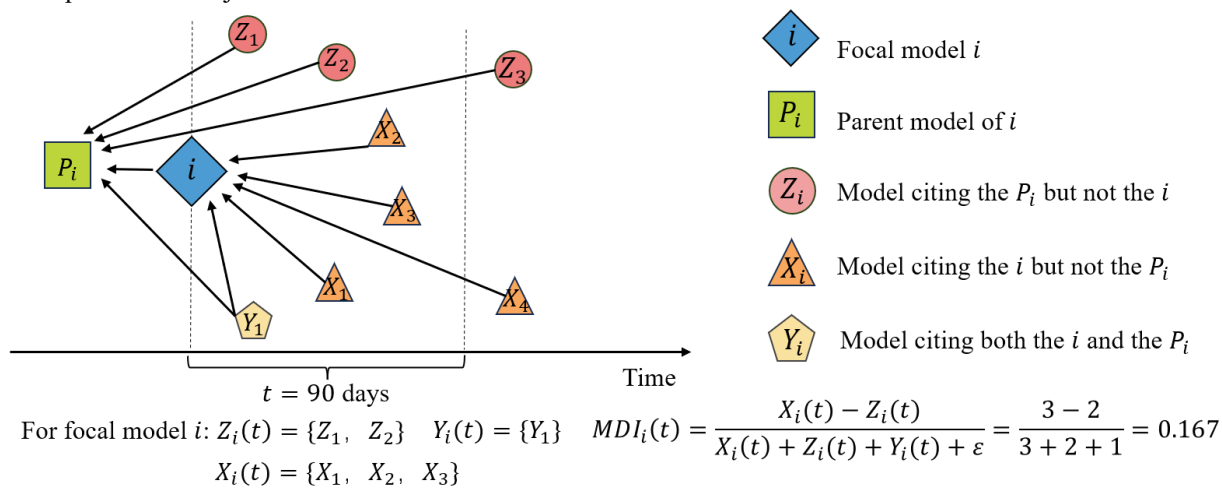

Figure 2. Illustration of the MDI Calculation.

## 4 RESULTS
### 4.1 Structural Characteristics of the LLM Lineage Network

As of February 3, 2026, we extracted a total of 750,495 models with valid derivation relationships. After removing anomalous edges, the resulting network comprises 685,139 nodes and 682,785 edges, indicating a highly sparse graph with an average degree of 1.99. Within this network, there are 61,386 base models, 243,502 finetuned models, 270,177 adapter models, 124,908 quantized models, and 14,455 merged models.

We first examined the structure of the lineage network (see Figure 3). As presented in Figure 3(a), the overall in-degree distribution exhibits a pronounced heavy-tailed pattern and is broadly consistent with a power-law distribution ($\alpha = 1.92$, $x_{\min} = 5$, goodness-of-fit $D = 0.0198$). Similar heavy-tailed tendencies are also observed across the major derivation types shown in Figure 3(a), suggesting that the concentration of downstream derivation is a general feature of the ecosystem rather than one restricted to a single relation category. This result indicates that while most models attract little or no downstream derivation activity, a small minority of models absorb a disproportionate share of lineage relationships.

This concentration is further reflected in the component structure of the network. We identified 29,362 weakly connected components (WCC) (see Figure 3(b)). The largest component contains 299,490 nodes and 324,020 edges, accounting for 43.70% of all nodes in the network. This giant component includes widely-reused base models such as google-bert/bert-base-uncased, Qwen/Qwen1.5-7B and openai-community/gpt2, suggesting that much of the open-source LLM ecosystem evolves around a relatively small set of backbone models. At the same time, the large number of small peripheral components indicates that many projects remain weakly connected to, or entirely detached from, the mainstream lineage network. Together, these results reveal a highly concentrated evolutionary structure characterized by strong path dependence, thereby motivating the subsequent analysis of whether some models nevertheless succeed in redirecting downstream development.

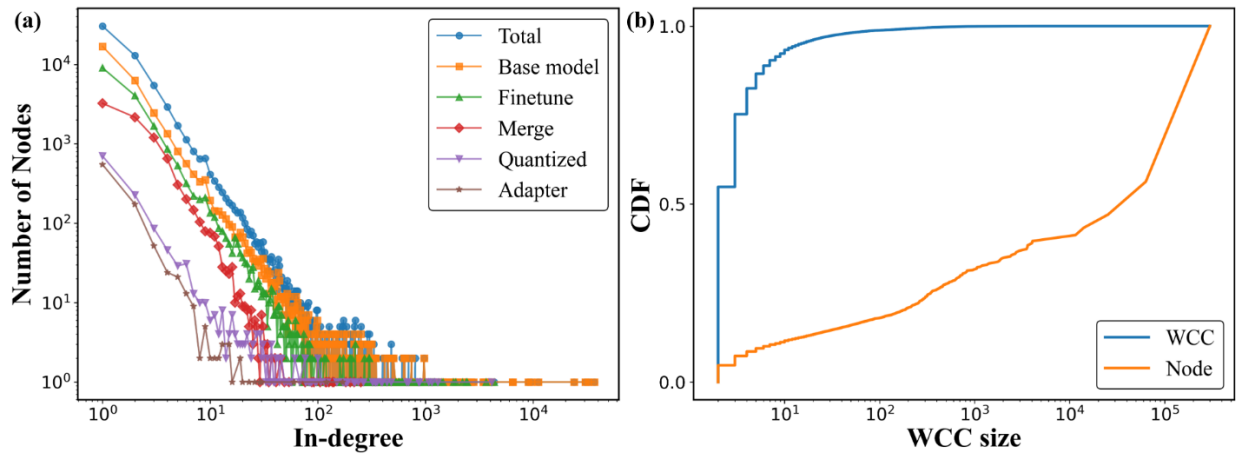

**Figure 3. Structural Characteristics of the LLM Lineage Network.** **(a)** In-degree distributions of the overall network and major derivation types, showing a strongly heavy-tailed pattern. **(b)** Cumulative distribution of weakly connected component (WCC) sizes, highlighting the coexistence of a giant component and many small peripheral components.

### 4.2 MDI Reveals a Predominantly Consolidative Community with Uneven Disruptive Potential

To ensure the interpretability of the MDI, we restricted the analysis to intermediate models that have both identifiable predecessor trajectories and observable downstream derivations. Accordingly, we excluded base models, which do not have parent models by definition, and terminal models, defined here as models that have no subsequent derivatives anywhere in the observed dataset. Models that had no descendants within a given observation window but did have later descendants elsewhere in the dataset were retained. This yielded 29,286 intermediate nodes for the MDI analysis. As presented in Figure 4(a), the overall distribution of MDI is strongly concentrated in the negative range, with a mean of -0.53. The most prominent peak occurs around MDI = -1.0, indicating that the vast majority of focal models are followed by downstream development that remains attached to their predecessor trajectories rather than being redirected toward the focal model itself. In other words, most models in the open-source LLM community function as consolidative rather than disruptive innovations. A small concentration appears around MDI = 0, suggesting a mixed response in which subsequent developers are divided between building on the focal model and continuing to rely on its parent trajectory. Finally, a small peak around MDI = 1.0 indicates that only a limited minority of models occupy clearly disruptive positions in the lineage network, becoming new focal points for subsequent derivation. Taken together, these results suggest that although genuinely disruptive models do exist, the dominant pattern in the community is one of consolidation under strong path dependence.

We further examined the association between model in-degree and MDI. As shown in Figure 4(b), we observed a positive correlation, which is supported by the Spearman correlation coefficient and further reflected in the upward trend of the LOWESS fit curve. In general, models with higher in-degrees are more likely to exhibit higher MDI values, suggesting that disruptive positions are more likely to emerge among models that attract substantial follow-on derivation. At the same time, positive MDI values remain rare at low and moderate in-degrees, indicating that the gravitational pull of predecessor trajectories is difficult to overcome unless a focal model achieves sufficiently broad downstream uptake. The LOWESS curve crosses the zero line at an in-degree of approximately 149, suggesting a descriptive turning point beyond which positive MDI values become more frequently observed. Together, these results indicate that disruptive downstream redirection is closely associated with a model's capacity to attract subsequent development at scale.

Furthermore, we examined how MDI varies across model parameter scales. For descriptive purposes, we grouped models into three categories: small-scale models (<1B), medium-scale models (1-10B), and large-scale models (>10B). We extracted the parameter scale size from model identifiers when available. For example, "14B" was identified from "Qwen/Qwen2.5-14B-Instruct". Among the 29,286 models with valid MDI values, parameter-scale information could be identified for 18,707 models. As presented in Figure 4(c), MDI values differ across parameter-scale categories. In descriptive terms, large-scale models exhibit a relatively higher MDI distribution than medium-scale and small-scale models. The group medians further reflect this pattern, increasing from -0.9636 for small-scale models to -0.9259 for medium-scale models and -0.8000 for large-scale models. This suggests that larger models are more likely to occupy less consolidative, and in some cases more disruptive, positions in the lineage network.

Finally, we further examined how MDI varies across different derivation strategies using the raincloud plots shown in Figure 4(d). The results suggest clear heterogeneity across relation types. Among the four major strategies, finetuned models exhibit the highest MDI distribution overall and are more likely to extend into the disruptive range (positive MDI values). Although most finetuned models still remain in the negative domain, this pattern suggests that finetuning is more likely than the other derivation strategies to generate downstream redirection away from predecessor trajectories. Quantization and model merging display comparable overall distributions, but both appear more polarized than finetuned or adapter models, with observations clustering more heavily toward the consolidative end while still retaining a smaller disruptive tail. By contrast, adapter models exhibit the weakest disruptive tendency overall. These differences should be interpreted cautiously, since the MDI reflects downstream lineage patterns rather than direct architectural or performance-based disruption.

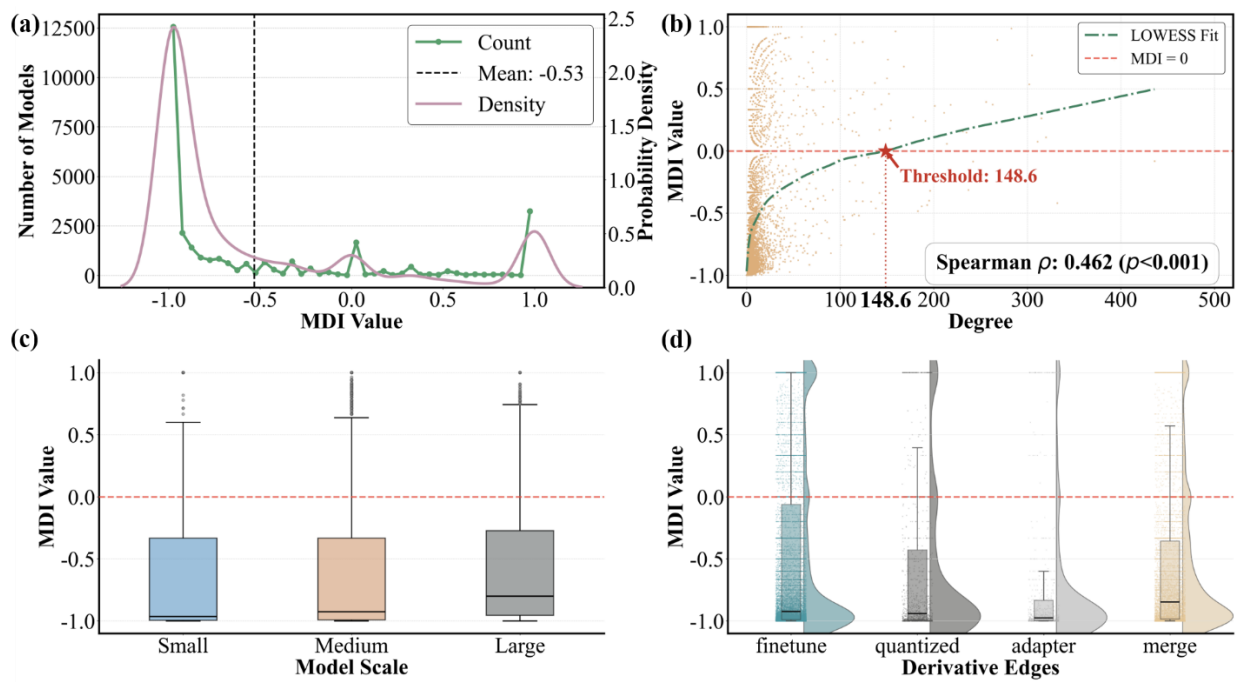

**Figure 4. Overall Pattern and Correlates of the Model Disruption Index (MDI). (a)** Overall distribution of MDI values across valid intermediate models. **(b)** Association between model in-degree and MDI, with a LOWESS fitted curve and the zero-disruption reference line. **(c)** Distribution of MDI values across broad parameter-scale categories, with the black line indicating the median. **(d)** Distribution of MDI values across major derivation strategies.

## 4.3 Consolidation Remains Dominant Over Time

To examine how disruption and consolidation evolve over time in the open-source LLM community, we analyzed the temporal pattern of the MDI from monthly, period-based, and observation-window perspectives. As shown in Figure 5(a), the number of monthly new models experienced exponential growth starting in early 2023 and has maintained a massive scale since then. Over the same period, the proportion of models with positive MDI values declined sharply after an initial early spike, but later exhibited a fluctuating upward trend. This pattern suggests that although disruptive downstream redirection did not disappear as the ecosystem expanded, it remained limited relative to the much larger volume of newly introduced models.

To provide a broader descriptive comparison of temporal change, we divided the timeline into four broad periods based on the release of major base models, namely the release of Llama 1 in March 2023, Llama 3 in April 2024, and Qwen 3 in April 2025 (see Figure 5(b)). In the earliest period, before March 2023, the MDI distribution shows a visible concentration near zero, suggesting uncertainty between building on the focal model and continuing along its parent trajectory. Between March 2023 and April 2024, the distribution becomes more polarized toward the two extremes. This polarization further intensifies between April 2024 and April 2025, and becomes most pronounced after April 2025. Taken together, these period-based comparisons suggest that, as the ecosystem matures, downstream development becomes less ambiguous in its response to new models: most derivatives continue to consolidate predecessor trajectories, while only a limited minority emerge as clearly disruptive alternatives.

We further examined the sensitivity of MDI to the length of the observation window. As presented in Figure 5(c), across observation windows ranging from 30 to 180 days, the overall pattern remains broadly consistent: the density of MDI values is concentrated in the consolidative range, while clearly disruptive models remain a small minority. At the same time, longer windows reduce the density around the neutral region and increase the mass near the strongly consolidative end. This suggests that, as more temporally distant downstream models are incorporated into the calculation, subsequent development is increasingly likely to remain attached to predecessor trajectories rather than continue to build on the focal model.

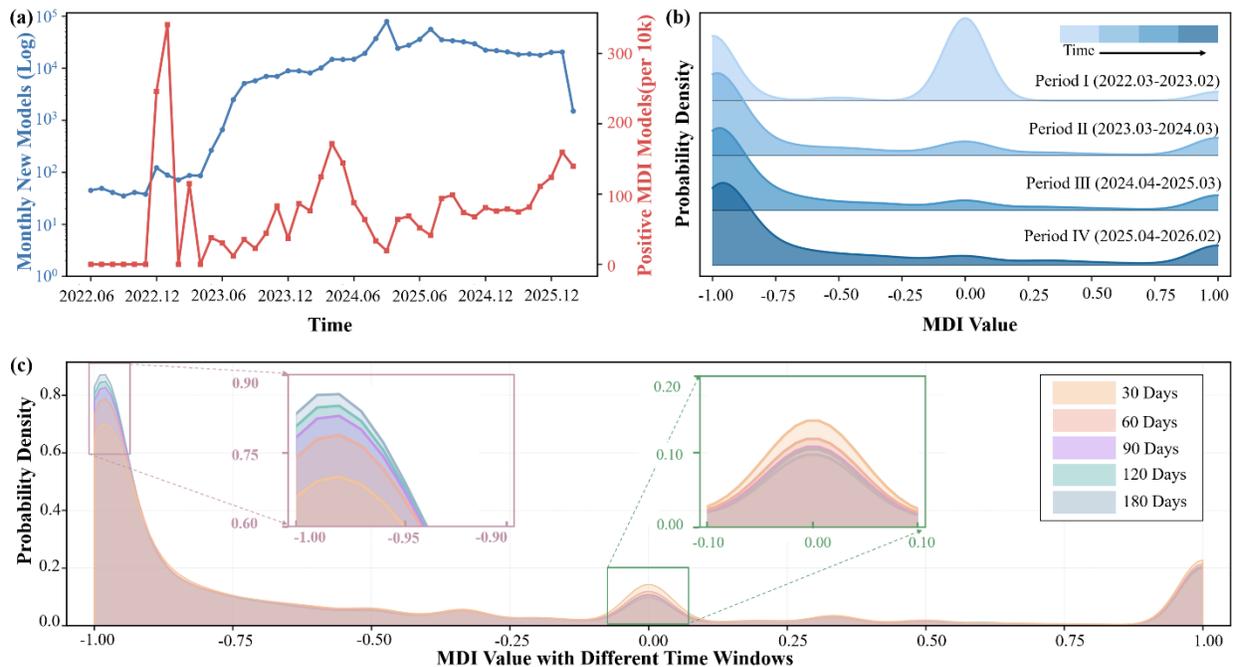

**Figure 5. Temporal Evolution of MDI. (a)** Monthly number of newly released models and the proportion of models with positive MDI values. **(b)** Ridge plots of MDI distributions across four broad periods anchored by major model

releases. **(c)** Sensitivity of MDI distributions to observation windows ranging from 30 to 180 days, showing stronger concentration in the consolidative range as the window length increases.

## 5 DISCUSSION

This study shows that the open-source LLM community on Hugging Face is not an innovation landscape in which downstream development is evenly distributed across models. In response to the questions raised at the outset of this study, our findings indicate that the community is structurally highly concentrated, that clearly disruptive models do emerge but remain a limited minority, and that downstream innovation accumulates unevenly around a relatively small set of dominant backbone models. Rather, the Hugging Face ecosystem is better understood as a lineage network characterized by strong path dependence, in which most models consolidate existing technological trajectories rather than redirecting subsequent development away from their predecessors. Taken together, these findings suggest that openness in model access does not necessarily translate into broadly distributed technological disruption.

At a theoretical level, these results connect open-source LLM research to a broader tradition in scientometrics and the study of technological change. The heavy-tailed, power-law-like concentration observed in the lineage network is consistent with a familiar finding in paper and patent studies: downstream attention and reuse are typically concentrated around a relatively small number of highly visible nodes (Redner, 1998; Valverde et al., 2007; Wu et al., 2019). Likewise, the MDI results suggest that the distinction between disruption and consolidation, originally developed for papers and patents, also captures an important feature of open-source model ecosystems. Although the objects differ, the underlying dynamic appears similar: only a limited minority redirect later activity, while most contributions extend and stabilize existing trajectories (Wu et al., 2019; Park et al., 2023).

This parallel with paper and patent systems is especially important for understanding how open-source LLM communities evolve. Existing work has often emphasized lineage tracing, provenance, and supply-chain dependencies in the LLM ecosystem (Bommasani et al., 2023; Shang et al., 2026; Wu et al., 2026). Our results suggest that the open-source community should also be understood as a socio-technical coordination system in which a small number of backbone models organize much of the downstream developmental space. In this sense, open access does not simply broaden participation; it also amplifies cumulative reuse around already visible and reusable models. The positive association between in-degree and MDI, together with the variation across parameter scales and derivation strategies, further suggests that disruptive downstream redirection is not evenly distributed across the community, but is more likely to emerge only under relatively restricted structural conditions. In descriptive terms, large-scale models tend to exhibit higher MDI values than smaller models, although this pattern should be interpreted cautiously and does not by itself establish a causal advantage of scale. This helps explain why rapid model proliferation in open-source LLMs does not necessarily imply an equally broad proliferation of new technological trajectories.

These findings should also be interpreted in light of several limitations, which point directly to future research. First, the MDI captures disruption in lineage structure rather than direct architectural, performance-based, or benchmark-based disruption. Second, the lineage graph depends on metadata disclosed on Hugging Face, and the completeness of parent-model documentation may vary across models and derivation practices. In particular, some technologically meaningful lineage relationships may not be represented as explicit edges on the platform. For example, transitions between major versions within the same model family, such as Qwen2 to Qwen3, may reflect clear evolutionary continuity without appearing as direct lineage links in Hugging Face metadata. Future work could address these limitations by combining lineage-based disruption with architectural analysis, benchmark performance, dataset dependencies, or developer behavior, and by incorporating additional lineage inference strategies to recover important but undocumented evolutionary connections.

## 6 CONCLUSION

In this study, we examined technological disruption in the open-source LLM community on Hugging Face from a scientometric network perspective. By constructing a large-scale lineage network and introducing the Model Disruption Index (MDI), we show that most models in the community function as consolidative rather than disruptive innovations, while only a limited minority emerge as new bases for subsequent development. These findings extend existing research on open-source LLM ecosystems by moving beyond descriptive lineage tracing to identify directional differences in downstream technological development. Future research could further improve this framework by incorporating undocumented lineage relations and additional evidence on model architecture, performance, and developer behavior.